\documentstyle[12pt]{article}

\begin{document}

\baselineskip=6.5mm

\renewcommand{\arraystretch}{1.3}
\newcommand{\be}{\begin{equation}}
\newcommand{\ee}{\end{equation}}
\newcommand{\ba}{\begin{array}}
\newcommand{\ea}{\end{array}}
\newcommand{\bea}{\begin{eqnarray}}
\newcommand{\eea}{\end{eqnarray}}
\newcommand{\refs}[1]{(\ref{#1})}
\newcommand{\ns}{\normalsize}

\begin{titlepage}
\title{{\Large\bf Quark and Lepton Mass Matrices 
                  from Horizontal U(1) Symmetry }\\
                          \vspace{-4.5cm}
                          \hfill{\ns CBNU-TH 960323\\}
                          \hfill{\ns SNUTP 96-023\\}
                          \hfill{\ns UPR-692T \\}
                          \hfill{\ns hep-ph/9605377\\}
                          \vspace{3.5cm} }

\author{Eung Jin Chun \\[.5cm]
  {\ns\it Department of Physics, Chungbuk National University}\\
  {\ns\it Cheongju, Chungbuk 360-763, Republic of Korea} \\[.5cm]
        Andr\'e Lukas\\[.5cm]
  {\ns\it Department of Physics, University of Pennsylvania}\\
  {\ns\it Philadelphia, PA 19104-6396, USA} \\[.5cm]
  }
\date{}
\maketitle
\begin{abstract} 
\baselineskip=7.2mm  {\ns
In the simplest model of horizontal U(1) symmetry with one singlet 
added to the supersymmetric standard model, we systematically reconstruct 
quark mass matrices from the low-energy data to prove that there are 
only two mass matrices found by Binetruy et.~al..  The same U(1) symmetry 
constrains the hierarchical structure of L-violating couplings, 
from which we build radiative neutrino mass matrices accommodating 
the solar and hot dark matter neutrino masses and mixing.  
We find a few patterns of acceptable charged lepton and neutrino 
mass matrices, most of which are consistent with large 
$\tan\beta \simeq m_t/m_b$ only.
}\end{abstract}

\thispagestyle{empty}
\end{titlepage}

\noindent {\bf 1. Introduction}
\medskip

Understanding the hierarchies of fermion masses is one of the fundamental
problems in particle physics. One way to explain the structure of 
fermion mass matrices is to assume a generation-dependent $U(1)$ symmetry, 
as originally explored in \cite{fn}.  
In this approach, the horizontal $U(1)$ symmetry constrains
non-renormalizable couplings of quarks and leptons to certain $SU(3)_c \times
SU(2)_L \times U(1)_Y$ singlet field. The hierarchical values of Yukawa 
couplings are given by some powers of the vacuum expectation value of the 
singlet spontaneously breaking the horizontal $U(1)$ symmetry.

Recent investigations \cite{lns}--\cite{blr} have been amounted to 
reveal some remarkable features of this class of  models.  
First of all, a potential connection to string theory was pointed out in 
\cite{ir,br,js}.  In the simplest model with only one
singlet field, phenomenologically allowed quark mass hierarchies require
the $U(1)$ to be anomalous.  Then the cancellation of the anomalies may be
the consequence of Green-Schwarz mechanism in string theories \cite{gs},
a remarkable feature of which is the prediction of the mixing
angle $\sin^2\theta_W = 3/8$ \cite{iba}. 
It was also shown that the horizontal $U(1)$ symmetry may provide
a solution to the Higgs mass problem (the $\mu$-problem)
\cite{nir,js2,chun}.  More recently, detailed analyses have been made to 
prove that this class of models indeed can account for the structure of 
the CKM mixing matrix \cite{dps,blr}.
In particular, for the simplest model with one singlet, only two sets of 
quark Yukawa matrices are found to be consistent with the experimental 
values \cite{blr}.
In the first half of this paper, we will systematically reconstruct the
quark Yukawa matrices starting from the experimental values to prove 
there are no other acceptable quark mass matrices.

Contrary to the quark sector, the lepton sector is not much constrained.
Our aim in the second half is to single out acceptable mass matrices for 
the charged leptons and neutrinos.  Even if there are not yet
decisive evidences for non-zero neutrino masses, some strong indications
{}from astronomical and cosmological observations have been widely discussed
\cite{smi}.  Theoretical efforts to understand neutrino mass structure in the
context of horizontal abelian symmetry were recently made by several groups 
\cite{dll,gn,blr}.  
They mostly rely on the see-saw mechanism to generate light neutrino 
masses. In this paper, we will look for  R-parity
violating terms (in particular, L-violating terms) which are inherent in the
minimal supersymmetric standard model (MSSM).  Without 
an additional symmetry like R-parity, the MSSM allows the presence of L- or 
B-violating operators;
$$ \Lambda LQD \,,\quad  \Lambda' LLE \,,\quad \Lambda'' UDD \,.  $$
Proton stability imposes an extremely strong constraint on $\Lambda\Lambda''$.
On the other hand, in the presence of the above L-violating operators
neutrino masses can be generated radiatively. In general, these masses
are not very much constrained due to the large freedom in the coupling
constants $\Lambda$ and $\Lambda'$. As we will see, this situation
changes significantly once a horizontal $U(1)$ symmetry determines the 
structure of the R-parity violating operator \cite{bn,bgnn}. Then, since
the charge assignment is fixed to a large extent due to the known masses
and anomaly cancellation the order of $\Lambda$, $\Lambda'$ and the radiative
neutrino mass matrices depends on a few charge values only. We will analyze
these matrices and show that a few solutions compatible with the solar
neutrino data and a hot dark matter neutrino mass exist. Simultaneously, the 
B-violating operators $UDD$ can be forbidden by the $U(1)$ symmetry to ensure
proton stability.
\bigskip

\noindent {\bf 2. Reconstruction of quark mass matrices}
\medskip

We consider the supersymmetric standard model whose Yukawa structure 
originates from horizontal abelian symmetry $U(1)_X$.  
Taking the simplest form,
we introduce only one singlet field $\phi$ whose $U(1)_X$-invariant
non-renormalizable couplings to quarks and leptons determine hierarchical
patterns of the Yukawa matrices.  
We assume that non-renormalizable couplings are
suppressed by the string scale (or Planck scale) $M$ invoking a string origin 
of the model.  The charges of $U(1)_X$ can be any integer.
When $\phi$ carries charge $N$, spontaneous breaking of $U(1)_X$
due to the vacuum expectation value $\langle \phi \rangle$ leaves $Z_N$ 
as an unbroken subgroup. For convenience we normalize the charge of $\phi$
to $-1$ and thus the charges of quarks, leptons and Higgses are 
integers divided by $N$.  
We assume that spontaneous breaking of $U(1)_X$ occurs slightly below
the scale $M$ giving rise to the Cabbibo angle $\epsilon \equiv
\langle \phi \rangle/M \simeq 0.22$ in order to predict the experimental 
values of the CKM mixing angles \cite{wol}.
\medskip

The large mass of top quark leads us to assume that the (3,3)-component of 
the up-type quark Yukawa couplings is given by the renormalizable term
$Q_3U_3H_2$ which implies $q_3+u_3+h_2=0$. We use the capital letters
$Q, U, D, L, E$ and $H_{1,2}$ to denote MSSM superfields,  and the
corresponding small letters denote their charges under $U(1)_X$.  
Depending on the value of $\tan\beta$, the bottom quark Yukawa
term can be written as $Q_3D_3H_1(\phi/M)^x$ where $x=q_3+d_3+h_1$ is a not 
too large positive integer.  Note that $\tan\beta\simeq \epsilon^x m_t/m_b$.
The excess charge matrices $y^u_{ij} \equiv q_{i3}+u_{j3}$ and 
$y^d_{ij} \equiv q_{i3}+d_{j3}$ (normalized by $y^{u,d}_{33}=0$)
 determine the Yukawa matrices for the up- and down-type quarks
\be  \label{yuk}
  Y^u_{ij} \sim \epsilon^{y^u_{ij}}\,\theta(y^u_{ij}) \,,\quad
  Y^d_{ij} \sim \epsilon^{x+y^d_{ij}}\,\theta(x+y^d_{ij}) 
\ee
where $q_{ij} \equiv q_i-q_j$, etc. and
the function $\theta$ is defined as  $\theta(x)=1$  when $x$ is a 
{\it non-negative integer}, and $\theta(x)=0$ otherwise.
As noted in \cite{dps,blr}, possible zeros in the Yukawa matrices \refs{yuk} 
are filled by  non-minimal contributions to the kinetic term
consistent with $U(1)_X$.  Their major role is to change the naive mixing 
angles calculated from the Yukawa matrices \refs{yuk}.  In the canonical basis
corresponding to normalized kinetic terms, the corrected Yukawa
matrices are \cite{dps}
\be \label{cyuk}
 \hat{Y}^u_{ij} \sim \sum_{l,m}\epsilon^{|q_{il}|+|u_{jm}|}\,Y^u_{lm}\,,\quad
 \hat{Y}^d_{ij} \sim  \sum_{l,m}\epsilon^{|q_{il}| + |d_{jm}|}\,Y^d_{lm} \,,
\ee 
where, e.g., $\epsilon^{|q_{il}|}$ vanishes when $|q_{il}|$ is
non-interger.

Taking the dominant contribution, one obtains for the down-type;
 $\hat{Y}^d_{ij}= \epsilon^{x+n^d_{ij}}$ where
\be \label{nij}
 n^d_{ij}= 
   \left\{ \ba{l}
    q_{i3}+d_{j3}\,, \quad \mbox{ when $q_{i3}+d_{j3}+x$ is 
                        a non-negative integer } \\
   {\rm Min}\,[ \hat{\theta}(|q_{il}|)+ \hat{\theta}(|d_{jm}|)+
                \hat{\theta}(q_{l3}+d_{m3}+x)]-x\,,  \quad
                   \mbox{otherwise} \,. \ea\right.
\ee
Here Min on the second line picks up the smallest value varying $l,m$ from 1 
to 3.  The function $\hat{\theta}$ assigns infinity when the argument is 
{\it negative or fractional} number and $\hat{\theta}(x)=x$ otherwise.
As observed in \cite{blr}, the expressions for $n_{i3}$ and $n_{3j}$ are 
simple: 
\be \label{i3}
  n^u_{i3}=n^d_{i3}=|q_{i3}|\,,\quad 
  n^u_{3j}=|u_{j3}|\, ,\quad n^d_{3j}=|d_{j3}| \,.
\ee 
We note that this property holds independently of the value of $x$ 
assuming that the (3,3)-component is the largest: $n_{33} \leq n_{ij}$. 
{}From the property $n_{ij} \geq |y_{ij}|$ \cite{blr} and
$y_{ij}=y_{i3}+y_{3j}$ one finds
\bea  \label{ineqs}
  n_{ij} &\leq& n_{i3} + n_{3j} \nonumber\\
  n_{i3} &\leq& n_{ij} + n_{3j}  \\
  n_{3j} &\leq& n_{ij} + n_{i3}  \nonumber 
\eea
These inequalities $n^{u,d}_{ij}$ allow us to completely reconstruct the
Yukawa matrices from the experimental values of the mass eigenvalues 
and the CKM matrix.  In refs.~\cite{dps,blr}, two mass matrices are
found to be acceptable.  In the below we will give a systematic proof of their
uniqueness.
\medskip

The eigenvalues of the Yukawa matrices $\hat{Y}$ determined by $n_{ij}$ in
\refs{nij} are  $\epsilon^{\rho_1}$,  $\epsilon^{\rho_2}$ and 1 where
\bea \label{rhos}
  \rho_1 &=& \mbox{Min}\,[n_{11}+n_{22},\, n_{12}+n_{21}]-\rho_2 \\
  \rho_2 &=& \mbox{Min}\,[n_{11}, n_{22},\, n_{12}, n_{21}] \nonumber
\eea
in case that $\rho_1>\rho_2$ \cite{br,blr}.
The experimental values of $\rho_{1,2}$ extrapolated to the Planck scale 
are  known to be $(\rho_1^u,\rho_2^u)=(8, 4)$ and 
$(\rho_1^d,\rho_2^d)=(4,2)$ for the up- and down-type quarks, 
respectively \cite{rrr}.  Notice that one can fix $n_{22}=\rho_2$ (namely, 
$n^u_{22}=4$ and $n^d_{22}=2$) by reordering $q_{i3}, u_{i3}$ and $d_{i3}$ and
by correct hierarchical values between the first and the second generations.
Then, one gets $n_{ij} \geq n_{22}$ for $i,j=1,2$. 
Another important property is that
\be \label{regular}
  y_{11}+y_{22} = y_{12}+y_{21} = \rho_1 + \rho_2
\ee
since the determinant of the Yukawa matrix \refs{yuk} should be regular and 
cannot change its value even after the kinetic correction.
The regularity condition of $Y_{ij}$ also shows that both of two elements 
in the same raw or column of the left-upper 2$\times$2 submatrix cannot 
be negative or fractional at the same time.  From these properties and 
eq.~\refs{rhos}, we arrive at the conclusion that e.g. the down-type excess 
charge matrices yielding $n^d_{22}=2$ is characterized by one of the 
following two patterns:
\bea\label{down}
  y^d_{11}=q_{13}+d_{13}=4 \,, && y^d_{22}=q_{23}+d_{23} =2 \,; \nonumber\\
  y^d_{11}= 8\,,\quad y^d_{22}=-2 && \mbox{with}\quad q_{23}=0, -1, -2 \,.
\eea
The latter condition comes from the fact that $n^d_{22}=|q_{23}|+|d_{23}|=2$
when $y_{22}^d<0$.
Similar conclusion can be drawn also for the up-type
quarks:  
\bea\label{up}
 q_{13}+u_{13}=8\,,\quad q_{23}+u_{23}=4 \,;\nonumber\\
 |q_{23}|+|u_{23}|=4 \quad\mbox{if}\quad q_{23}+u_{23}<0 \,.
\eea
The above properties are useful to reconstruct quark mass matrices compatible
with the CKM mixing angles.
\medskip

The CKM matrix in the leading order can be parameterized in terms of three
left-handed mixing angles as follows:
\be  \label{ckm}
  V_{CKM} = \left( \ba{ccc} 
        1 & -S^-_{12}-S^u_{13}S^-_{23} & -S^-_{13}+S^u_{12}S^-_{23}\\
        S^-_{12}+S^d_{13}S^-_{23} & 1 & -S^-_{23}-S^u_{12}S^-_{13}\\
        S^-_{13}-S^d_{12}S^-_{23} & S^-_{23}+S^d_{12}S^-_{13} & 1 \ea \right)
\ee
where $S^-_{ij}=S^u_{ij}-S^d_{ij}$ \cite{hr}.  
With the assumption $Y_{ij} \leq Y_{33}$, an order of magnitude calculation
can be made to express the mixing angles $S_{ij}$ in terms of the Yukawa 
elements \cite{hr}.  In the model under consideration, 
the expressions become extremely simple.
Iterative use of \refs{ineqs} reveals
\bea \label{sij}
  s_{13} &=& n_{13}\,, \quad s_{23} = n_{23} \\
  s_{12} &=& \mbox{Min}\,[n_{12}-n_{22},\, n_{11}+n_{21}-2n_{22}] \nonumber
\eea
where $S_{ij} \sim \epsilon^{s_{ij}}$.
{}From \refs{i3}, we immediately get $s^u_{i3}=s^d_{i3}=|q_{i3}|$.
Comparing the CKM matrix \refs{ckm} to the experimental data \cite{wol}, we 
find two possibilities:
\bea \label{justwo}
  s^u_{12} \mbox{ or } s^d_{12}=1\,,\quad 
 &s^u_{13} \mbox{ or } s^d_{13}=3\,,\quad &
  s^u_{23} \mbox{ or } s^d_{23}=2 \,;\nonumber\\  
  s^u_{12} \mbox{ and } s^d_{12}=1\,,\quad 
 &s^u_{13} \mbox{ and } s^d_{13}>3\,,\quad &
  s^u_{23} \mbox{ or } s^d_{23}=2   \,.
\eea 
The above conditions can be rewritten as
\bea \label{justwo'}
  |q_{23}|=2\,,&\quad |q_{13}|=3\,,\quad& s^u_{12}\mbox{  or  }s^d_{12}=1\,; 
        \nonumber\\
  |q_{23}|=2\,, &\quad |q_{13}|>3\,,& \quad s^u_{12}=s^d_{12}=1 \,.
\eea
It is now useful to realize that the condition $s_{12}=1$ cannot be met by the 
term $s_{12}=n_{11}+n_{21}-2n_{22}$ since $n_{11}\geq 2n_{22}$ 
(see \refs{regular}) and $n_{21}\geq n_{22}=\rho_2$. 
Therefore, one gets $n_{12}=s_{12}+\rho_2$ for $s_{12}=1$.
After some manipulation just with the down-type quark mass matrices satisfying
\refs{down}, one finds that the second condition in \refs{justwo'} cannot be
fulfilled.  In order to examine the first case in \refs{justwo'}, one 
tries out mass matrices satisfying $(q_{13},q_{23})=(\pm 3, \pm2)$ 
under the conditions in \refs{down} and \refs{up}.  
Then we find that $s^u_{12}$ or $s^d_{12}=1$ can only be obtained for 
$(q_{13}, q_{23})=(3,2)$ and $(-3,2)$.
These corresponds to the two mass matrices found in \cite{blr}.
The first one has the excess charges 
$(q_{13}, q_{23})=(3,2)$, $(u_{13}, u_{23})=(5,2)$ and 
$(d_{13}, d_{23})=(1,0)$ producing the Yukawa matrix
\be \label{yuk1}
  \mbox{(I)} \qquad\quad
  \hat{Y}^u \sim \left( \ba{ccc}
                   \epsilon^8 & \epsilon^5 & \epsilon^3 \\
                   \epsilon^7 & \epsilon^4 & \epsilon^2 \\
                   \epsilon^5 & \epsilon^2 & 1 \ea\right) \,,\qquad
  \hat{Y}^d \sim \epsilon^x \left( \ba{ccc}
                   \epsilon^4 & \epsilon^3 & \epsilon^3 \\
                   \epsilon^3 & \epsilon^2 & \epsilon^2 \\
                   \epsilon^1 & 1 & 1 \ea\right) \,.
\ee
For the second one,  $(q_{13}, q_{23})=(-3,2)$, $(u_{13}, u_{23})=(11,2)$ and 
$(d_{13}, d_{23})=(7,0)$; the resulting Yukawa matrix is 
\be \label{yuk2}
  \mbox{(II)} \qquad\quad
  \hat{Y}^u \sim \left( \ba{ccc}
                   \epsilon^8 & \epsilon^5 & \epsilon^3 \\
                   \epsilon^{13} & \epsilon^4 & \epsilon^2 \\
                   \epsilon^{11} & \epsilon^2 & 1 \ea\right)\,, \qquad
  \hat{Y}^d \sim \epsilon^x \left( \ba{ccc}
                   \epsilon^4 & \epsilon^3 & \epsilon^3 \\
                   \epsilon^9 & \epsilon^2 & \epsilon^2 \\
                   \epsilon^7 & 1 & 1 \ea\right) \,.
\ee
We also remark  that the experimental values of the mass eigenvalues 
and the CKM mixing angles cannot be reproduced when any of $y_{ij}$ is 
fractional.
\bigskip

\noindent{\bf 3. Lepton mass matrices and R-parity violating couplings}
\medskip

The lepton sector is only weakly constrained compared to the quark sector.
Furthermore, the MSSM allows L- and B-violating terms whose
presence may destroy proton stability. The usual wisdom is to 
assign R-parity to forbid both or either of them.  As one can see,  the
horizontal $U(1)$ symmetry compatible with the experimental quark and lepton
mass matrices still has large number of free charges. As a result, 
it is always possible to choose some fractional or negative excess 
charges so that undesirable R-parity violating operators can be discarded.
One can also give large enough positive excess charges to suppress the proton 
decay.
In the below we will investigate the case where L-violating 
operators are allowed but B-violating operators are sufficiently
suppressed. The structure of the L-violating operators as well as
the radiatively generated neutrino masses resulting from the $U(1)_X$ symmetry
will be analyzed.

The measurement of solar neutrino deficit strongly indicates the 
neutrino mass and mixing;  
\be
  m_{\nu_\alpha} \simeq 2\times 10^{-3} \mbox{  eV},\quad 
  \theta_{e\alpha} \simeq 3 \times 10^{-2} \sim \epsilon^2
\ee
which provides the right amount of resonance conversion $\nu_e \to \nu_\alpha$
($\alpha = \mu, \tau$) inside the Sun \cite{smi}.  
Other hints for neutrino masses come from the deficit of
atmospheric neutrinos and the need for hot dark matter in structure
formation.
Given hierarchical Yukawa matrices, it appears unnatural to accommodate all of 
three evidences which require nearly degenerate neutrino mass matrices 
\cite{cm}.
Therefore we will analyze to what extent the $U(1)$-constrained radiative
neutrino mass matrices can provide the solar neutrino mass and mixing as well
as a hot dark matter neutrino mass $m_{\nu} \sim 10$ eV.
\medskip

For the charged leptons, one gets the following Yukawa matrix elements: 
\be
  Y^e_{ij} \sim \epsilon^{x+y^e_{ij}}\,\theta(x+y^e_{ij})
  \qquad\mbox{with}\quad y^e_{ij} = l_{i3} + e_{j3} 
\ee
where we have used the $b$--$\tau$ unification at the Planck scale,
$x=l_3+e_3+h_1=q_3+d_3+h_1$.
The  couplings  of the R-parity violating terms are 
\bea
  \Lambda_{ijk} &\sim& \epsilon^{y_{ijk}}\,\theta(y_{ijk})\qquad
               \mbox{with}\quad y_{ijk}= l_0+l_{i3}+q_{j3}+d_{k3} \nonumber\\
  \Lambda'_{ijk} &\sim& \epsilon^{y'_{ijk}}\,\theta(y'_{ijk})\qquad
               \mbox{with}\quad y'_{ijk}= l_0+l_{i3}+l_{j3}+e_{k3} \nonumber\\
  \Lambda''_{ijk} &\sim& \epsilon^{y''_{ijk}}\,\theta(y''_{ijk})\qquad
               \mbox{with}\quad y''_{ijk}= b_0+u_{i3}+d_{j3}+d_{k3} \nonumber
\eea
where $l_0 \equiv l_3+q_3+d_3=l_3+l_3+e_3$ and $b_0 \equiv u_3+d_3+d_3$.
One has to bear in mind that $\Lambda'_{iij} =\Lambda''_{kll} \equiv 0$ due to
the antisymmetric contractions. 
In the similar way as in the down quark sector, the experimental
eigenvalues of the charged lepton mass matrix; $\rho_1=4, \rho_2=2$
\cite{br} eliminate two free charges; 
\bea \label{elec}
 &e_{13}=4 - l_{13}\,,& e_{23}=2-l_{23}\,; \nonumber\\
 \mbox{or}\quad &e_{23}=-2 -l_{23}& \quad \mbox{with}\quad l_{23}=-2,-1,0 \,.
\eea
From now on, we will concentrate on the first possiblitiy as it turns
out that the second one doesnot yield proper lepton mass matrices for
our purpose.
The number of free charges is further reduced by considering
the anomaly-free conditions of $U(1)_X$.  
The anomalies corresponding to $SU(3)^2-U(1)_X$, 
$SU(2)^2-U(1)_X$, $U(1)_Y^2-U(1)_X$ and $U(1)_Y-U(1)_X^2$ are 
\bea
  A_3 &=& \sum_i(2 q_i + u_i + d_i) \nonumber\\
  A_2 &=& \sum_i (3 q_i + l_i) + (h_1+h_2)  \\
  A_1 &=& \sum_i ({1\over3}q_i+{8\over3}u_i+{2\over3}d_i+l_i+2 e_i)+(h_1+h_2)
          \nonumber\\
  A'_1 &=& \sum_i(q_i^2-2u_i^2+d_i^2-l_i^2+e_i^2)- (h_1^2-h_2^2) \nonumber \,.
\eea
The desirable quark mass eigenvalues satisfying $\sum_i(q_{i3}+u_{i3}) =12$,
$\sum_i(q_{i3}+d_{i3})=\sum_i(l_{i3}+e_{i3})=6$ is not consistent with 
$A_3=A_2=A_1=0$ \cite{br}.  But the MSSM with horizontal $U(1)_X$ symmetry
may come from superstring theory which allows the Green-Schwarz anomaly 
cancellation mechanism \cite{gs} ; $k_3g_3^2=k_2g_2^2=k_1g_1^2 \propto
g^2_{string}$. Here $k_{i}$ are Kac-Moody levels of the gauge groups which are
integers for non-abelian groups.  For the anomaly cancellation,
the ratio of the anomalies should follow $A_3:A_2:A_1=k_3:k_2:k_1$.
The gauge coupling unification in the MSSM still holds quite well 
even near the Planck scale, which implies
$k_3=k_2=3k_1/5=1$ \cite{iba}.  Therefore, the Green-Schwarz mechanism
indicates the relations among the anomalies:  $A_3:A_2:A_1=1:1:5/3$.

Then the relation $\sum_i(q_{i3}+d_{i3})=\sum_i(l_{i3}+e_{i3})$ results in 
$h_1+h_2=0$ allowing the presence of the Higgs mass term $\mu H_1H_2$ 
in the superpotential.  In this case, we are facing with the $\mu$-problem.
In the context of horizontal abelian symmetry, one may have a natural solution
by assuming a slight deviation of the hierarchical eigenvalues like 
$\rho^e_1=5$ \cite{nir} or when R-symmetry is responsible for the hierarchy
\cite{chun}.
The conditions $A_3=A_2=3A_1/5$ together with $A'_1=0$ leaves three
charges free which are chosen to be $l_0$, $l_{13}$ and $l_{23}$.
The other charges are given by
\bea \label{other}
  u_3 &=&-q_3+h_1\,,\quad d_3=x-q_3-h_1\,,\quad e_3=x-l_3-h_1\quad 
           l_3=l_0-x+h_1\,, \nonumber\\[3mm]
  q_3 &=&  2+{2\over 5}x - {1\over5}(b_{0}+1)
           -{1\over3}(q_{13}+q_{23})\,,\quad
  h_1 = {4\over5}x-{2\over5}(b_0+1) \\
  b_0 &=& {1\over3}(l_{13}+l_{23}+3l_0)-1 \nonumber 
\eea
considering the first case in \refs{elec}.
Now the couplings $\Lambda$, $\Lambda'$ and $\Lambda''$ are functions of $l_0$
and $l_{i3}$.  

The radiative neutrino masses (assuming minimal soft terms) due to
the couplings $\Lambda$ and $\Lambda'$ are 
\be
 m^\nu_{ij} \simeq {\mu\tan\beta+A \over 8\pi^2 m_0^2}
            \mbox{Tr}\,[ 3(\Lambda_i)M_d^T(\Lambda_j)M_d^T  +
                 (\Lambda'_i)M_e^T(\Lambda'_j)M_e^T ]\, , 
\ee
where $(\Lambda_i)_{jk} \equiv \Lambda_{ijk}$, etc.~and 
$A$ is the soft-SUSY breaking parameter of the trilinear terms and 
$m_0$ is the soft mass of squarks or sleptons.
Taking the leading order terms inside the trace, one can pick out the integers
$p_{ij}$ and $p'_{ij}$ as follows;
\bea \label{mn1}
 m^\nu_{ij}&\sim&{m_\tau^2 \over 8 \pi^2 m_0} 
  [3\left(m_b \over m_\tau \right)^2 \epsilon^{p_{ij}} + \epsilon^{p'_{ij}} ]
         \nonumber\\
  \mbox{with}\quad p_{ij}&=&\mbox{Min}\,[y_{ilm}+y_{jkn}+y^d_{ln}+y^d_{km}] \\
  \mbox{and}\quad p'_{ij}&=&\mbox{Min}\,[y'_{ilm}+y'_{jkn}+y^e_{ln}+y^e_{km}]
       \nonumber
\eea
where $\mu \simeq A \simeq m_0 \simeq 0.1-1$ TeV.  It is understood 
to neglect negative or fractional $y$'s inside Min.  

The expression \refs{mn1} is valid  at the Planck scale where 
$m_b/m_\tau \simeq 1$.  Renormalizing $M_d$, $M_e$
as well as $\Lambda$, $\Lambda'$ down to the weak scale, one gets the 
significant effect; $\Lambda/\Lambda' \simeq 3$ like $m_b/m_\tau \simeq 3$.  
Therefore the neutrino mass matrix at the weak scale is given by 
\be
 m^\nu_{ij} \sim \tilde{m} [ \epsilon^{p_{ij}-3} + \epsilon^{p'_{ij}}] 
\ee
where $\tilde{m} \simeq 0.4-0.04 \mbox{MeV}$.
\medskip

Writing $m^\nu_{ij} \sim \tilde{m} \epsilon^{n^\nu_{ij}}$, 
it is simple to get the number $n^\nu_{ij}$ in the case where 
$l_0$ and $l_0+l_{i3}$ are all positive integers.  
It is  convenient to calculate the neutrino mass matrix in 
the original basis with non-canonical kinetic term. 
As $l_0$,  $l_0+l_{i3}$ are positive integers, $\Lambda_{i33}$ is present 
and gives the largest contribution to the neutrino mass matrix so that 
\be
 n^\nu = 2l_0+ \left( \ba{ccc} 2l_{13} & l_{13}+l_{23} & l_{13} \\
                              l_{13}+l_{23} & 2l_{23} & l_{23} \\
                                l_{13} & l_{23} & 0 \ea\right)
         -3 \,.
\ee
When $\tilde{m} \simeq 0.04-0.4$ MeV, one can extract the following best 
values for $l_{i3}$, $l_0$ and $b_0$;
\be \label{res1}
l_{13}=5\,,\quad l_{23}=3\,,\quad l_0= 4 \,(5)\,,\quad 
 b_0= {17\over3} \, ({21\over3})
\ee
which (for $\tilde{m} \simeq 0.4$ MeV) give $m_{\nu_\tau} \sim 20 (10)$ eV,
$m_{\nu_\mu} \sim 2 (1)\times 10^{-3}$ eV and 
$\theta_{e\mu} \sim \epsilon^2$.  Even though the mass of a hot dark
matter neutrino is not precisely fixed, 
one hardly can allow one unit of deviation in $l_0$ since it leads 
to two order deviation ($\epsilon^2$) in the neutrino masses 
which is too far from the desired value. 
The same is also true for the solar neutrino mass.
The resulting neutrino mass matrices are
\be \label{res1n}
  m^\nu \sim \left(\ba{ccc} \epsilon^{15} & \epsilon^{13} & \epsilon^{10} \\
              \epsilon^{13} & \epsilon^{11} & \epsilon^{8} \\ 
              \epsilon^{10} & \epsilon^{8} & \epsilon^{5} \ea\right) 
          (\epsilon^{2}) \,.
\ee
We have chosen tau-neutrino as the heaviest one.
In either case in \refs{res1},  
the excess charge matrix and mass matrix in the canonical
basis for charged leptons are given by  
\be \label{res1e}
 y^e= \left(\ba{ccc} 4&4&5\\2&2&3\\-1&-1&0\ea\right) \,,\qquad
 \hat{M}^e \sim \left(\ba{ccc} \epsilon^4 & \epsilon^4 & \epsilon^5 \\
            \epsilon^2 & \epsilon^2 & \epsilon^3 \\ 
            \epsilon^1 & \epsilon^1 & 1\ea\right)\,.
\ee
It is interesting to note that the charged lepton mass matrix in \refs{res1e}
allows {\it only} $x=0$, that is, large $\tan\beta \simeq m_t/m_b$.
This conclusion holds for either matrix (I) or (II) in \refs{yuk1} and 
\refs{yuk2}. Furthermore, due to the fractional value of $b_0$ the
B-violating operators $UDD$ are forbidden. We also notice that
higher dimensional operators like $QQQL$, $uude$ dangerous for  
proton stability are also forbidden by the same reason.
\medskip

Now let us generalize the above calculation. 
Given any fractional numbers $l_0$, $l_{i3}$, one of the following cases 
may come out.
First, some of them can be fractional leading to degenerate matrices with
a $2\times 2$ nontrivial submatrix only. Matrices of this type cannot lead
to the correct solar neutrino and hot dark matter masses simultaneously and
are therefore disregarded. Second, $\Lambda$ and $\Lambda'$ are
allowed to generate proper neutrino masses. Then the operators $UDD$ should be
completely forbidden or highly suppressed to achieve proton stability;
$\Lambda \Lambda'' < 10^{-26}$. A large suppression turns out to be 
inconsistent with the desired neutrino masses due to the
last relation in \refs{other}. Therefore we have to choose $b_0$ fractional or
smaller than $-\mbox{Max}[u_i+d_j+d_k]$ which forbids $UDD$ completely.
As a third case one might try to allow $\Lambda'$ and 
$\Lambda''$ and forbid $\Lambda$ (or make it sufficiently small) which
would also ensure proton stability. As we will see, the wanted neutrino mass
matrix does not permit this possibility. 

To examine the above cases explicitly, we carry out a
systematic calculation of neutrino mass matrices as follows.
We start by calculating neutrino mass matrices in the original basis
as in \refs{mn1}.  Then we go to the canonical basis by performing 
a rotation $K_l$ for the lepton doublets $L_i$;
\be 
 K_l = \left( \ba{ccc} 1 & \epsilon^{|l_{12}|} & \epsilon^{|l_{13}|}\\
                  \epsilon^{|l_{12}|} & 1 & \epsilon^{|l_{23}|} \\
                  \epsilon^{|l_{13}|} & \epsilon^{|l_{23}|} & 1 \ea\right)\,.
\ee
Next step is to rotate the lepton doublet fields to diagonalize the 
charged-lepton mass matrix.  This can be done by the rotation matrix $R_l$;
\bea
 &R_l& = \left(\ba{ccc} 1&-s^e_{12}&0\\s^e_{12}&1&0\\0&0&1 \ea\right) 
       \left(\ba{ccc} 1&0&-s^e_{13}\\0&1&0\\s^e_{13}&0&1 \ea\right) 
       \left(\ba{ccc} 1&0&0\\0&1&-s^e_{23}\\0&s^e_{13}&1 \ea\right) \\
&\mbox{with}& s^e_{i3}=n^e_{i3} \,, \quad s^e_{12}=n^e_{12}-2 \nonumber
\eea
as can be seen from \refs{sij}.
Taking the leading order, the total rotation matrix $R=R_l K_l$ is given by
$R_{ij} \sim \epsilon^{r_{ij}}$ where $r_{ii}=0$ and 
\bea
 r_{12}&=&r_{21}=\mbox{Min}\,[n^e_{12}-2,\, |l_{12}|] \nonumber\\
 r_{13}&=&\mbox{Min}\,[n^e_{12}-2+|l_{23}|,\, |l_{13}|] \nonumber\\
 r_{23}&=&\mbox{Min}\,[n^e_{12}-2+|l_{13}|,\, |l_{23}|] \\
 r_{31}&=&|l_{13}|\,,\quad  r_{32}=|l_{23}| \nonumber \,.
\eea

In most cases of interest $\Lambda$ gives the dominant contribution for the
radiative neutrino mass. For case (I) in \refs{yuk1},
one finds no more acceptable neutrino mass matrices
even for some negative values of $l_0, l_{i3}$. To see this, let us first note 
that the resultant neutrino mass matrix is only two-by-two if one of $l_0$,
$l_0+l_{i3}$ is smaller than $-$Max$[q_{i3}+d_{j3}]$ ($= -4$ in case (I)).
Hence, $l_0, l_0+l_{i3} \geq -{\rm Max}[q_{i3}+d_{i3}]$ has to be met.
As one can see the smallest value ($\epsilon^5$) in the neutrino mass matrix
results from $l_0, \,(l_0+l_{i3}) =-4$. For this choice, the same radiative
neutrino masses and mixing as in \refs{res1n} can be obtained taking 
$l_{13}=13$, $l_{23}=11$ and $l_0=-4$.  
However, this leads to $b_0=3$ and therefore to a
too large $\Lambda''$. On the other hand, in case (II), we find one more
neutrino mass matrix, which  has negative integers  $l_0$ and $l_0+l_{23}$.
When the tau neutrino is the heaviest, the explicit pattern is given by
\be
l_{13}=19\,,\quad l_{23}=-3\,,\quad 
l_0 = -4\, (-3)\,,\quad b_0={1\over3}\, ({4\over3}) 
\ee
for $\tilde{m}= 0.04-0.4$ MeV.  The resultant neutrino mass matrix is
\be \label{res2n}
  m^\nu \sim \left(\ba{ccc} \epsilon^{27} & \epsilon^{13} & \epsilon^{16} \\
              \epsilon^{13} & \epsilon^{11} & \epsilon^{8} \\ 
              \epsilon^{16} & \epsilon^{8} & \epsilon^{5} \ea\right) 
         (\epsilon^{2}) \,.
\ee
As the corresponding charged lepton mass matrix in the canonical basis
we obtain
\be \label{res2e}
 y^e= \left(\ba{ccc} 4&24&19\\-18&2&-3\\-15&5&0\ea\right)\,, \qquad
 \hat{M}^e \sim \epsilon^x 
   \left(\ba{ccc} \epsilon^4 & \epsilon^{24} & \epsilon^{19} \\
                  \epsilon^{18} & \epsilon^{2} & \epsilon^{3} \\
                  \epsilon^{15} & \epsilon^{5} & 1 \ea\right)\,.
\ee
Notice that this charge matrix can be consistent with $x=0,1,2$ allowing also
small $\tan\beta$.
\medskip

The same calculation can be done with the muon neutrino as the heaviest.
For both cases (I) and (II), the values $(l_{13}, l_{23}) = (2,-3)$, 
$l_0= 7\,(8)$ and $b_0={17\over3}\,({20\over3})$ give rise to the following
neutrino mass matrix in the $M^e$-diagonal canonical basis,
\be \label{res3n}
 m^\nu \sim \left(\ba{ccc} 
                  \epsilon^{15} & \epsilon^{10} & \epsilon^{13} \\
                  \epsilon^{10} & \epsilon^{5} & \epsilon^{8} \\
                  \epsilon^{13} & \epsilon^{8} & \epsilon^{11} \ea\right)
                  (\epsilon^2) \,.
\ee
It differs from \refs{res1n} just by an exchange of the second and the third
generation. In the charged lepton sector this solution is characterized by
\be \label{res3e}
 y^e= \left(\ba{ccc} 4&7&2\\-1&2&-3\\2&5&0\ea\right)\,, \qquad
 \hat{M}^e \sim \epsilon^x 
   \left(\ba{ccc} \epsilon^4 & \epsilon^{7} & \epsilon^{2} \\
                  \epsilon^{5} & \epsilon^{2} & \epsilon^{3} \\
                  \epsilon^{2} & \epsilon^{5} & 1 \ea\right) \,.
\ee
For case (II), we find one more solution 
with $(l_{13},l_{23})=(22,3)$, $l_0=-7\,(-6)$ and $b_0={1\over3}\,({4\over3})$
yielding the neutrino mass matrix
\be \label{res4n}
 m^\nu \sim \left(\ba{ccc} 
                  \epsilon^{27} & \epsilon^{16} & \epsilon^{13} \\
                  \epsilon^{16} & \epsilon^{5} & \epsilon^{8} \\
                  \epsilon^{13} & \epsilon^{8} & \epsilon^{11} \ea\right)
                  (\epsilon^2) \,,
\ee
and the excess charge and mass matrix of charged leptons
\be \label{res4e}
 y^e= \left(\ba{ccc} 4&21&22\\-15&2&3\\-18&-1&0\ea\right)\,, \qquad
 \hat{M}^e \sim \epsilon^x 
   \left(\ba{ccc} \epsilon^4 & \epsilon^{7} & \epsilon^{12} \\
                  \epsilon^{19} & \epsilon^{2} & \epsilon^{3} \\
                  \epsilon^{18} & \epsilon^{1} & 1 \ea\right) \,.
\ee
When the muon neutrino is the heaviest, only $x=0$ is allowed as can be 
seen from \refs{res3e} and \refs{res4e}.
\bigskip

\noindent{\bf 4. Conclusion}
\medskip

We examined the model of horizontal $U(1)$ symmetry with  one singlet added to
MSSM, which explains the desired hierarchies of quarks and leptons.
This model turns out to be quite constrained by the experimental values of 
quark masses and mixing.  We systematically reconstructed the quark mass 
matrices to prove that there are only two acceptable patterns of quark Yukawa 
structures.

We extended the analysis to the charged leptons and neutrinos.
Combined use of the experimental  values of the charged lepton masses and 
Green-Schwarz mechanism of the anomaly cancellation of $U(1)$ reduces  many
free charges to three ($l_{13}, l_{23}, l_0$) in the lepton 
sector.  Depending on their values, the presence of L- or B-violating 
operators can be controlled.  Concentrating on L-violating operators (while
suppressing B-violating operators for proton stability) we singled out
a few patterns of the charged lepton and neutrino mass matrices which
accommodate the solar and hot dark matter neutrino masses and mixing.
Interestingly, only large $\tan\beta=m_t/m_b$ ($x=0$) is allowed when
the muon neutrino is the heaviest.  When the tau-neutrino is the heaviest, the
quark mass matrix (I) allows large $\tan\beta$ and only the pattern (II) is 
consistent with both large and small $\tan\beta$ ($x=0,1,2$).
\bigskip

\noindent{\bf Acknowledgments}
\medskip

One of the authors (E.J.C.) thanks J.E. Kim and E. Dudas for communications 
and discussions. 
A.L. thanks J.E. Kim and K. Choi for hospitality during his
stay in Korea where part of this work was done. This work is supported in part
by KOSEF-DFG interchange program and by CTP at Seoul National University.
E.J.C. is a Brain-Pool fellow.

\end{document}